\documentclass[
  prl,
  twocolumn,
  superscriptaddress,
  aps,
  amsmath,
  amssymb,
  floatfix,
  10pt,
  longbibliography
]{revtex4-2}

\usepackage{graphicx}
\usepackage{bm}
\usepackage{mathtools}
\usepackage{microtype}
\usepackage{xcolor}
\usepackage[colorlinks=true,allcolors=blue]{hyperref}

\newcommand{\ii}{\mathrm{i}}
\newcommand{\ee}{\mathrm{e}}
\newcommand{\US}{U_{SS}}
\newcommand{\Heff}{H_{\mathrm{eff}}}
\newcommand{\Tr}{\operatorname{Tr}}
\newcommand{\appref}[1]{#1}
\newcommand{\paperheading}[1]{\par\indent\phantomsection{\color{red}\itshape #1.}---\ignorespaces}
\newcounter{appref}
\renewcommand{\theappref}{\Alph{appref}}
\newcommand{\appsec}[2]{\par\refstepcounter{appref}\label{#1}\setcounter{equation}{0}\renewcommand{\theequation}{\theappref\arabic{equation}}\renewcommand{\theHequation}{\theappref.\arabic{equation}}\indent{\color{red}\itshape Appendix~\theappref: #2.}---\ignorespaces}

\begin{document}

\title{Non-Bloch Dynamical Quantum Phase Transitions with Memory}

\author{Zhesen Yang}
\email[Corresponding author: ]{yangzs@xmu.edu.cn}
\affiliation{Asia Pacific Center for Theoretical Physics, Pohang, Republic of Korea}
\affiliation{Department of Physics, Xiamen University, Xiamen 361005, Fujian Province, China}

\author{Wei-Wei Yang}
\email[Corresponding author: ]{yangww18@lzu.edu.cn}
\affiliation{Asia Pacific Center for Theoretical Physics, Pohang, Republic of Korea}
\affiliation{Beijing National Laboratory for Condensed Matter Physics and Institute of Physics, Chinese Academy of Sciences, Beijing 100190, China}

\author{Henning Schomerus}
\email[Corresponding author: ]{h.schomerus@lancaster.ac.uk}
\affiliation{School of Physics and Astronomy, Lancaster University, Lancaster, LA1 4YB, United Kingdom}

\author{Jiangping Hu}
\email[Corresponding author: ]{jphu@iphy.ac.cn}
\affiliation{Beijing National Laboratory for Condensed Matter Physics and Institute of Physics, Chinese Academy of Sciences, Beijing 100190, China}
\affiliation{New Cornerstone Science Laboratory, Beijing, 100190, China}

\begin{abstract}
  Dynamical quantum phase transitions (DQPTs) of periodic chains are determined by real Bloch momenta.
  We identify non-Bloch DQPTs, which occur only in open chains and are determined by complex momenta instead.
  They arise in a Hermitian chain coupled to a simple substrate, whose memory is treated exactly, and originate from the point-gap winding number $W$ of the projected propagator: for $W\neq0$, open boundaries change both the dynamical free energy and the critical times, even in the thermodynamic limit.
  We find that both follow from an analytic continuation to complex momenta, and show that $W$ can be measured by particle counting.
  This reveals a dynamical bulk-boundary correspondence, in which non-Hermitian topology emerges from Hermitian dynamics with memory.
\end{abstract}

\maketitle

\paperheading{Introduction}\label{sec:introduction}
In equilibrium systems with short-range interactions, boundaries generically change the free energy only by a surface term that vanishes per site in the thermodynamic limit~\cite{FisherLebowitz1970}.
Therefore, equilibrium phase transitions are bulk properties and can be studied under periodic boundary conditions (PBC).
Dynamical quantum phase transitions (DQPTs) extend phase transitions to real-time evolution, where a dynamical free energy becomes nonanalytic at critical times~\cite{Heyl2013,Heyl2018,Andraschko2014,Jurcevic2017}.
They have been studied for sudden quenches~\cite{Karrasch2013Quench,Vajna2014Disentangling,Kriel2014ANNNI,Heyl2015Scaling,Zunkovic2018LongRange} and slow ramps~\cite{Pollmann2010Sweep,Sharma2016SlowQuench,Divakaran2016XY,Bhattacharya2017RampedHaldane,Zamani2024RampedQuenchDQPT}, as well as in periodically driven systems~\cite{Yang2019FloquetDQPT,Jafari2021FloquetEntanglement,Zhou2021FloquetDQPTQuenched,Jafari2022Synchronized,Naji2022EngineeringFloquetDQPT}, topological systems~\cite{Vajna2015,BudichHeyl2016,Huang2016TopologicalNodes,Flaschner2018,Sedlmayr2018}, and disordered systems~\cite{Yin2018AndersonZeros,Gurarie2019RandomField,Mendl2019Disorder,Cao2020WeakDisorder,Trapin2021RandomIsing}.
By analogy with equilibrium, one expects boundaries to affect neither the dynamical free energy nor the critical times, although exceptions have been reported~\cite{Khatun2019,Kaur2026}.

Conventional DQPTs depend on the initial state~\cite{Heyl2018,Zvyagin2016,Andraschko2014,Xu2024QuantumCoherenceDQPT}; subspace DQPTs generalize them by replacing this state with a subspace of a closed system~\cite{YangDQPTI}.
At a subspace DQPT, the dynamical free energy becomes nonanalytic, and the effective non-Hermitian Hamiltonian $\Heff(t)=\ii[\partial_t\US(t)]\US^{-1}(t)$ of the subspace diverges~\cite{YangDQPTI}. This Hamiltonian is time dependent, because the environment induces non-Markovian effects.
Since non-Hermitian systems can be sensitive to boundaries, as in the non-Hermitian skin effect (NHSE)~\cite{Yao2018,Zhang2020Winding,Okuma2020,Zhang2022UniversalNHSE,Zhang2022NHSEReview,OkumaSato2023Review,Lin2023NHSEReview,Gohsrich2025Perspective}, subspace DQPTs may also depend on boundaries.

\begin{figure}[!t]
  \centering
  \includegraphics[width=0.88\columnwidth]{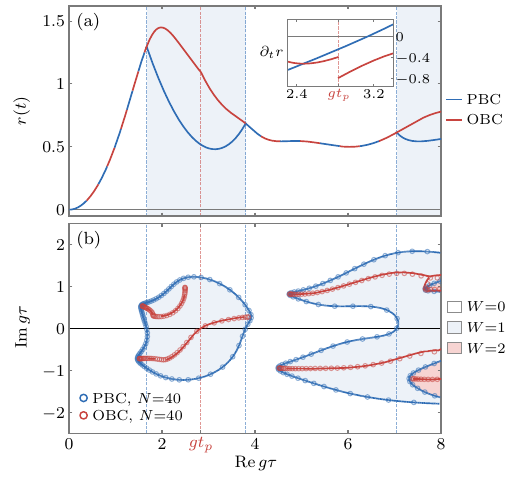}
  \caption{Boundary-dependent subspace DQPTs.\label{fig:obc}
    (a)~Dynamical free energy $r(t)$ under PBC (blue) and OBC (red, from non-Bloch band theory) in the thermodynamic limit; where the two coincide, the curve alternates in color. Inset: $\partial_t r$ (in units of the chain-substrate coupling $g$) near the non-Bloch DQPT at $gt_p=2.83$, where the OBC slope jumps.
    (b)~Fisher zeros, i.e., complex times $\tau$ with $\det\US(\tau)=0$, in the thermodynamic limit (curves) and for $N=40$ cells (circles), colored as in (a); black line: real-time axis of (a).
    Shading in (a) and (b): winding number $W$ (key on the right); dashed lines: PBC (blue) and non-Bloch (red) DQPTs. Numerical details: \appref{Appendix~\ref{app:model}}.}
\end{figure}

In this Letter, we study the subspace DQPTs of a Hermitian chain coupled to a simple substrate, whose memory makes the dynamics of the chain non-Markovian.
We find that the dynamical free energy of an open chain, i.e., under open boundary conditions (OBC), differs from that under PBC in certain time intervals, even in the thermodynamic limit [Fig.~\ref{fig:obc}(a)].
This difference has a bulk topological origin, namely the point-gap topology~\cite{Kawabata2019} of $\US(k,\tau)$, the time-evolution operator projected onto the chain at momentum $k$ in the Brillouin zone (BZ) and time $\tau$, which we also allow to be complex. This topology is characterized by the winding number~\cite{Gong2018}
\begin{equation}\label{eq:Wintro}
  W(\tau)=\oint_{\mathrm{BZ}}\frac{\mathrm dk}{2\pi}\,\partial_k\arg\det\US(k,\tau).
\end{equation}
The OBC and PBC dynamical free energies differ whenever the real-time axis passes through a region with $W\neq0$ [shaded in Fig.~\ref{fig:obc}(b)].
The PBC DQPTs occur where $W$ changes along the real-time axis.
The OBC and PBC DQPTs differ completely, both in their critical times and in the momenta that determine them, which are real for PBC and complex for OBC. Under OBC, all PBC DQPTs disappear [Fig.~\ref{fig:obc}(a)], and a nonzero $W$ can induce a new DQPT at a time $t_p$, where the OBC Fisher zeros [red in Fig.~\ref{fig:obc}(b)] cross the real-time axis.
Guided by the non-Bloch band theory of non-Hermitian systems~\cite{Yao2018,Yokomizo2019,Yokomizo2020Review,WangAmoeba2024}, we find that the OBC dynamical free energy [red curve in Fig.~\ref{fig:obc}(a)] and $t_p$ in the thermodynamic limit follow from the PBC propagator continued to complex momenta. We therefore call this transition a non-Bloch DQPT.
Finally, building on full counting statistics~\cite{Levitov1996,Esposito2009} and site-resolved fermion detection~\cite{Cheuk2015,Parsons2015}, we propose a scheme to measure $W$ by particle counting in fermionic many-body systems.

\paperheading{Subspace DQPTs}\label{sec:dqpt_comparison}
We first review the definition of subspace DQPTs.
Consider a subspace $S$ and its orthogonal complement $E$ in the Hilbert space of a closed system, whose Hermitian Hamiltonian takes the block form
\begin{equation}\label{eq:Hblocks}
  H=\begin{pmatrix}
    H_S & H_{SE} \\
    H_{ES} & H_E
  \end{pmatrix},\qquad H_{ES}=H_{SE}^{\dagger},
\end{equation}
where $H_S$ ($H_E$) acts within $S$ ($E$), and $H_{SE}$ couples them.
The subspace dynamical amplitude is defined as~\cite{YangDQPTI}
\begin{equation}\label{eq:subdqpt}
  \mathcal D(t)=\det\US(t)=\det\bigl[\Pi_S\ee^{-\ii Ht}\Pi_S^\dagger\bigr],
\end{equation}
where $\Pi_S$ maps the full system onto $S$, so that $\Pi_S\Pi_S^\dagger=1$ on $S$ and $\Pi_S^\dagger\Pi_S$ is the projector onto $S$, and $\US(t)$ is the projected propagator.
Numerically, we compute $\US(t)=\sum_n\Pi_S|n\rangle\ee^{-\ii\omega_nt}\langle n|\Pi_S^\dagger$ for a finite system from the eigenvalues $\omega_n$ and eigenstates $|n\rangle$ of $H$.
Since amplitude leaks from $S$ into $E$, $\US(t)$ is not unitary, and its eigenvalues lie in the unit disk.
When $S$ consists of $N$ unit cells, the dynamical free energy is defined as
\begin{equation}\label{eq:subdqpt_rate}
  r(t)=-\frac1N\ln|\mathcal D(t)|.
\end{equation}
A subspace DQPT occurs when the Fisher zeros~\cite{Fisher1965,Bena2005,Heyl2018}, i.e., the zeros of $\mathcal D$ in the complex-time plane, approach the real axis in the thermodynamic limit~\cite{Heyl2013,Andraschko2014,YangDQPTI},
\begin{equation}\label{eq:subdqpt_condition}
  \mathcal D(\tau)=\det\US(\tau)=0,\quad
  \tau\to t_c\in\mathbb R\ \ (N\to\infty),
\end{equation}
leading to a nonanalytic $r(t)$ at the critical time $t_c$.

\begin{figure}[t]
  \centering
  \includegraphics[width=0.9\columnwidth]{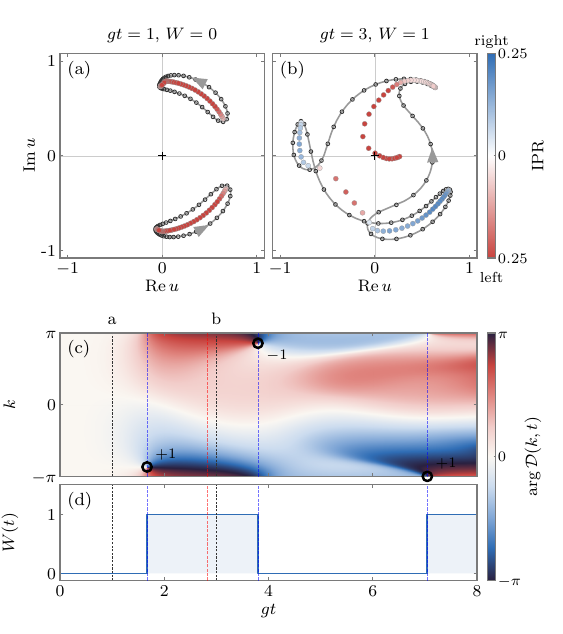}
  \caption{Non-Markovian skin effect and dynamical vortices.\label{fig:phase}\label{fig:spec}
    (a),(b) Eigenvalues $u$ of $\US$ at $gt=1$ and $3$ under PBC for continuous $k$ (gray curves; arrows: increasing $k$) and for $N=40$ cells under PBC (open circles) and OBC (dots); $+$ marks $u=0$. At $gt=3$, the two eigenvalues are interchanged after one turn of $k$ and together encircle $u=0$ once.
    Dot color: red (blue) for right eigenvectors centered in the left (right) half of the chain; intensity: their inverse participation ratio.
    (c)~$\arg\mathcal D(k,t)$; circles: vortices labeled by their charges $q$.
    (d)~Winding number $W(t)$.
    Dashed lines: the times of (a) and (b) (labels a, b), the PBC DQPTs (blue), and the non-Bloch DQPT (red). Numerical details: \appref{Appendix~\ref{app:model}}.}
\end{figure}

\paperheading{Model}\label{sec:model}
As a concrete example, we take $S$ to be a Su-Schrieffer-Heeger (SSH) chain with broken time-reversal symmetry, coupled to the edge of a semi-infinite two-dimensional substrate $E$ \appref{(Appendix~\ref{app:model})}.
Under PBC, $\US$ is block diagonal in the momentum $k$, so that $\mathcal D(t)=\prod_k\mathcal D(k,t)$ with $\mathcal D(k,t)=\det\US(k,t)$.
Eliminating the substrate gives, for $t>0$,
\begin{equation}\label{eq:Dk}
  \US(k,t)=\ii\int\frac{\mathrm d\omega}{2\pi}\,\ee^{-\ii\omega t}\bigl[\omega+\ii0-H_S(k)-\Sigma(k,\omega)\bigr]^{-1},
\end{equation}
where $\omega$ is the frequency, the chain Hamiltonian reads
\begin{equation}\label{eq:hS}
  H_S(k)=(t_1+t_2\cos k)\sigma_x+t_2\sin k\,\sigma_y+2\lambda\sin k\,\sigma_z,
\end{equation}
and the retarded self-energy, derived in \appref{Appendix~\ref{app:model}}, reads
\begin{equation}\label{eq:sigma}
  \Sigma(k,\omega)=\frac{g^2}{2J_y^2}\Bigl[\omega-\varepsilon_k-\sqrt{(\omega-\varepsilon_k)^2-4J_y^2}\Bigr]|B\rangle\langle B|.
\end{equation}
Here, the Pauli matrices act on the sublattices $A$ and $B$, $t_1$ ($t_2$) is the intracell (intercell) hopping, and the next-nearest-neighbor hopping $\lambda$ breaks time-reversal symmetry.
The coupling $g$ connects $B$ to the substrate edge, $-J_x$ ($-J_y$) is the substrate hopping along (across) the chain, $\varepsilon_k=-2J_x\cos k$, and the square root takes its retarded branch.
Throughout this work, we use $t_1=g$, $t_2=0.5g$, $\lambda=0.3g$, $J_x=0.15g$, and $J_y=0.25g$, with $g$ setting the energy scale.

\paperheading{Non-Markovian skin effect and winding number}\label{sec:winding}
As shown in Figs.~\ref{fig:spec}(a) and \ref{fig:spec}(b), the eigenvalues $u$ of $\US$ under PBC and OBC differ, and the OBC eigenstates accumulate near the chain ends, i.e., $\US$ shows an NHSE.
If $\US(k,t)$ were of the Markovian form $\ee^{-\ii H_{\rm M}(k)t}$ with a time-independent $H_{\rm M}$, its eigenstates, and hence this NHSE, would not change with time.
Since $\Sigma$ depends on $\omega$, $\US$ is not of this form, and its NHSE changes with time: at $gt=1$, all OBC eigenstates accumulate at the left end [Fig.~\ref{fig:spec}(a)], whereas at $gt=3$, they accumulate at both ends [Fig.~\ref{fig:spec}(b)].

The dynamical free energy is fully determined by the eigenvalues $u_n$ of $\US$, $r=-N^{-1}\sum_n\ln|u_n|$. Surprisingly, however, the NHSE does not necessarily lead to different dynamical free energies under PBC and OBC.
At $gt=1$, for instance, the PBC and OBC eigenvalues clearly differ [Fig.~\ref{fig:spec}(a)], yet the two dynamical free energies coincide [Fig.~\ref{fig:obc}(a)].

Numerically, we find that the two dynamical free energies differ only when $W$ becomes nonzero [Figs.~\ref{fig:obc}(a) and \ref{fig:phase}(d)], i.e., when the PBC spectrum winds around $u=0$, as at $gt=3$ [Fig.~\ref{fig:spec}(b)] but not at $gt=1$ [Fig.~\ref{fig:spec}(a)]; we explain this below.
The winding number $W$ can change only when $\mathcal D(k,t)$ vanishes, i.e., at a vortex around which $\arg\mathcal D$ winds by $2\pi q$~\cite{BudichHeyl2016,Flaschner2018}, with the charge $q=\pm1$ counted counterclockwise in the $(t,k)$ plane of Fig.~\ref{fig:phase}(c).
By Stokes' theorem, $W$ jumps by the total charge of the vortices at that time.
Since $\mathcal D(t)=\prod_k\mathcal D(k,t)$, the PBC Fisher zeros reach the real axis exactly at the times of the vortices, which thus mark the PBC DQPTs.
For our parameters and $gt\le8$, they lie at $gt_c=1.67$, $3.80$, and $7.05$, each with a single vortex of charge $+1$, $-1$, and $+1$, respectively [Figs.~\ref{fig:phase}(c) and \ref{fig:phase}(d)].

\paperheading{Dynamical free energy of open chains}\label{sec:nonbloch}\label{sec:ronkin}
We now explain why $r_{\rm OBC}\neq r_{\rm PBC}$ in the intervals with $W=1$ [Fig.~\ref{fig:obc}(a)].
In the thermodynamic limit, the sum over $k$ of $\ln|\mathcal D(k,t)|$ becomes an integral, and the PBC dynamical free energy is given by
\begin{equation}\label{eq:rate}
 r_{\rm PBC}(t)=-\int_{-\pi}^{\pi}\frac{\mathrm dk}{2\pi}\ln|\mathcal D(k,t)|.
\end{equation}
We continue the momentum analytically, $k\to k-\ii\mu$ with real $\mu$. This moves $z=\ee^{\ii k}$ from the unit circle to the circle $|z|=\ee^{\mu}$.
For each $\mu$, the continued PBC dynamical free energy reads
\begin{equation}\label{eq:ronkin}
 r_{\rm PBC}(\mu,t)=-\int_{-\pi}^{\pi}\frac{\mathrm dk}{2\pi}\ln|\mathcal D(k-\ii\mu,t)|,
\end{equation}
which reduces to $r_{\rm PBC}(t)$ at $\mu=0$.

\begin{figure}[!b]
  \centering
  \includegraphics[width=0.9\columnwidth]{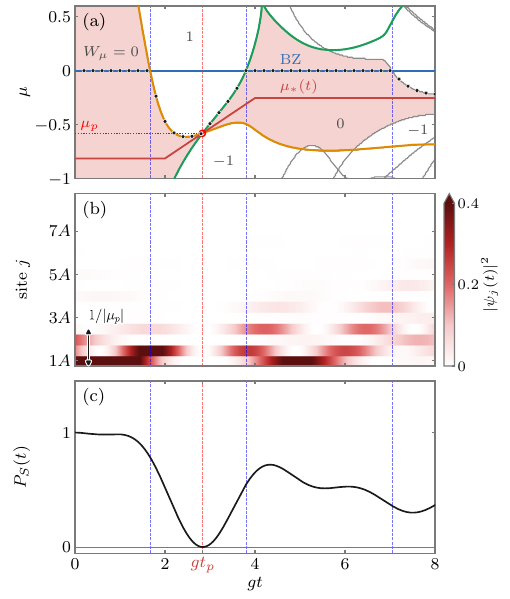}
  \caption{\label{fig:fisher}%
    Analytic continuation and non-Bloch DQPT.
    (a)~Winding number $W_\mu$ (numbers) in the $(t,\mu)$ plane, with winding boundaries (gray lines, except the two that bound the plateau in the first interval with $W=1$, in orange and green), the zero-winding plateau (shading) closing at $(t_p,\mu_p)$ (red circle), the BZ (blue line), and the path $\mu_*(t)$ (red curve); dots: $r_{\rm PBC}(t)-r_{\rm OBC}(t)$ at intervals of $0.2/g$.
    (b)~Density $|\psi_j(t)|^2$ of $\ee^{-\ii Ht}|r\rangle$ on the first 8 cells of an open chain with $N=40$, where $|r\rangle$ is the right eigenvector of $\US(t_p)$ with the eigenvalue closest to zero; arrow: localization length $1/|\mu_p|$.
    (c)~Survival probability $P_S(t)$ of $|r\rangle$ on the chain.
    Dashed lines as in Fig.~\ref{fig:obc}; numerical details: \appref{Appendix~\ref{app:model}}.}
\end{figure}

Guided by non-Bloch band theory~\cite{Zirnstein2021,Xue2021,WangAmoeba2024}, we find our main result: under OBC, 
\begin{equation}\label{eq:robc}
 r_{\rm OBC}(t)=r_{\rm PBC}(\mu_*(t),t),
\end{equation}
where $\mu_*(t)$ is any $\mu$ with zero winding number,
\begin{equation}\label{eq:winding_mu}
 W_\mu(t)=\int_{-\pi}^{\pi}\frac{\mathrm dk}{2\pi}\,\partial_k\arg\mathcal D(k-\ii\mu,t)=0.
\end{equation}
These $\mu$ form the zero-winding plateau [shaded in Fig.~\ref{fig:fisher}(a)]. The plateau is bounded by the winding boundaries, across which $W_\mu$ jumps [lines in Fig.~\ref{fig:fisher}(a)], and on it, $r_{\rm PBC}(\mu,t)=r_{\rm OBC}(t)$.
Evaluating $r_{\rm PBC}(\mu,t)$ along the BZ ($\mu=0$, blue line) and along the path $\mu_*(t)$ (red curve) in Fig.~\ref{fig:fisher}(a) gives the blue and red curves in Fig.~\ref{fig:obc}(a), respectively. The red curve agrees with finite-size numerics for open chains \appref{(Appendix~\ref{app:nonbloch})}.
Plotting their difference $r_{\rm PBC}(t)-r_{\rm OBC}(t)$ as dots in Fig.~\ref{fig:fisher}(a), we find that the dots lie on the BZ when $W=0$ and on the winding boundary just below the BZ when $W=1$; in the first interval with $W=1$, they lie first on the orange and then on the green one.
To explain this, we use the Cauchy--Riemann relation $\partial_\mu\ln|\mathcal D|=\partial_k\arg\mathcal D$ at the complex momentum $k-\ii\mu$, which turns Eqs.~\eqref{eq:ronkin} and \eqref{eq:winding_mu} into
\begin{equation}\label{eq:max}
\partial_\mu r_{\rm PBC}(\mu,t)=-W_\mu(t).
\end{equation}
For $W=0$, the BZ lies on the plateau, so that $r_{\rm OBC}=r_{\rm PBC}$. For $W=1$, $W_\mu=1$ between the upper edge $\mu_+(t)$ of the plateau, i.e., the winding boundary just below the BZ, and the BZ, so that integrating Eq.~\eqref{eq:max} from the plateau up to the BZ gives
\begin{equation}\label{eq:tp}
r_{\rm OBC}(t)-r_{\rm PBC}(t)=\int_{\mu_+(t)}^{0}W_\mu\,\mathrm d\mu=-\mu_+(t),
\end{equation}
the distance from the BZ down to this winding boundary.

\begin{figure*}[t]
  \centering
  \includegraphics[width=\textwidth]{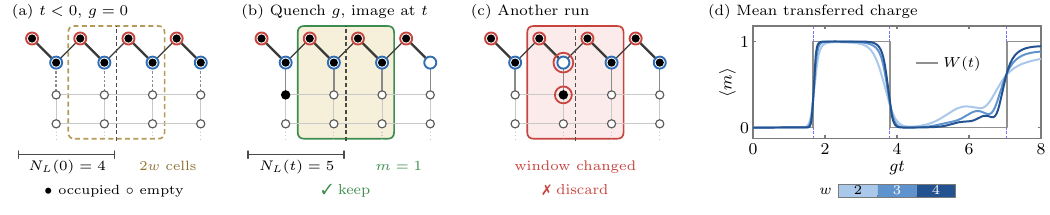}
  \caption{\label{fig:fcs}%
    Many-body detection of the winding number.
    (a)--(c)~Protocol. A filled chain and an empty substrate (a) evolve after $g$ is switched on. A snapshot taken at time $t$ is kept if its window of $2w$ cells returns to the initial pattern (b) and discarded otherwise (c); for each kept snapshot, the transferred charge $m=N_L(t)-N_L(0)$ is recorded.
    Black disks mark occupied sites, and red (blue) circles the $A$ ($B$) sites; in (c), large red rings mark the sites that differ from the initial pattern.
    (d)~Conditional mean $\langle m\rangle$ for $w=2,3,4$, compared with $W(t)$; dashed lines: PBC DQPTs.
    Open chain with $N=30$ cells \appref{(Sec.~\ref{sm:method} of the Supplemental Material)}.}
\end{figure*}

\paperheading{Non-Bloch DQPT}\label{sec:pinch}
The winding boundaries in Fig.~\ref{fig:fisher}(a) locate both kinds of DQPTs.
The PBC DQPTs (blue dashed lines) occur where a winding boundary crosses the BZ, e.g., the orange one at $gt=1.67$.
The non-Bloch DQPT occurs instead where the orange and green winding boundaries cross and the zero-winding plateau closes, at $gt_p=2.83$ and $\mu_p=-0.58$ [red circle in Fig.~\ref{fig:fisher}(a)].
To see why $r_{\rm OBC}$ has a kink there, recall from Eq.~\eqref{eq:tp} that, for $W=1$, $r_{\rm OBC}$ exceeds $r_{\rm PBC}$ by the distance $|\mu_+(t)|$ from the BZ down to the upper edge of the plateau [dots in Fig.~\ref{fig:fisher}(a)].
Before $t_p$, $\mu_+(t)$ follows the orange winding boundary, and after $t_p$ the green one, so that its slope changes abruptly at $t_p$.
Since $r_{\rm PBC}$ is smooth there, $r_{\rm OBC}$ has a kink [inset of Fig.~\ref{fig:obc}(a)].
At the PBC DQPTs, on the other hand, the kink of $r_{\rm PBC}$ is canceled by that of $r_{\rm OBC}-r_{\rm PBC}$, so that $r_{\rm OBC}$ is smooth.
The size of each kink follows from the slopes of the winding boundaries \appref{(Appendix~\ref{app:nonbloch})}, e.g., $-2.02g$ for $r_{\rm PBC}$ at $gt=1.67$ and $-(0.545-0.146)g=-0.399g$ for $r_{\rm OBC}$ at $t_p$, with the slopes $0.146g$ and $0.545g$ of the orange and green winding boundaries.
The closing is not accidental: the orange winding boundary opens the interval with $W=1$, and the green one closes it, so that $\mu_+(t)$ must switch from orange to green in between, which can happen only where the two edges of the plateau meet, i.e., where the plateau closes.

The value of $\mu_p$ has two physical meanings.
First, since $\mu_+(t_p)=\mu_p$, Eq.~\eqref{eq:tp} gives the gap between the red and blue curves in Fig.~\ref{fig:obc}(a) at the transition, $r_{\rm OBC}(t_p)-r_{\rm PBC}(t_p)=-\mu_p=0.581$.
Second, $\mu_p$ sets the size of the state that escapes from the chain at $t_p$.
The OBC eigenvalue $u$ of $\US(t_p)$ closest to zero vanishes in the thermodynamic limit ($|u|=0.024$ for $N=40$), and its normalized right eigenvector $|r\rangle$ decays as $\ee^{\mu_pj}$, i.e., with the localization length $1/|\mu_p|=1.7$ cells [Fig.~\ref{fig:fisher}(b)].
Since $\US(t_p)|r\rangle=u|r\rangle$, the probability $P_S(t)=\|\US(t)|r\rangle\|^2$ that this state stays on the chain equals $|u|^2\to0$ at $t_p$, and, since $\mathcal D=\prod_nu_n$, an OBC Fisher zero reaches the real axis [Fig.~\ref{fig:obc}(b)].
For $N=40$, the state almost completely leaves the chain at $t_p$ ($P_S=6\times10^{-4}$) and then partly returns [Figs.~\ref{fig:fisher}(b) and \ref{fig:fisher}(c)], a signature of memory: under a Markovian evolution, $P_S$ could only decrease.

\paperheading{Many-body extension and detection}\label{sec:manybody}
For noninteracting spinless fermions, subspace DQPTs are conventional DQPTs of a many-body quench from the state with every chain site occupied and the substrate empty,
\begin{equation}\label{eq:psi0}
 |\Psi_0\rangle=\prod_{i\in S}c_i^\dagger|0\rangle,
\end{equation}
where $i\in S$ runs over the sites of the chain.
At $t=0$, we switch on $g$ and measure the return probability, or Loschmidt echo, $|\mathcal G(t)|^2$ under $\hat H=\sum_{ij}c_i^\dagger H_{ij}c_j$~\cite{Heyl2013,Jurcevic2017}.
Since the evolution preserves the Slater-determinant form, the Loschmidt amplitude equals the subspace dynamical amplitude in Eq.~\eqref{eq:subdqpt}~\cite{Lowdin1955,Klich2003},
\begin{equation}\label{eq:echo}
 \mathcal G(t)=\langle\Psi_0|\ee^{-\ii\hat Ht}|\Psi_0\rangle=\det\US(t).
\end{equation}
Therefore, the dynamical free energy and the subspace DQPTs, including their boundary dependence, can in principle be detected through the Loschmidt echo.

Besides the dynamical free energy and the DQPTs, we find that $W(t)$ itself can be measured by particle counting, which requires only an open chain [Figs.~\ref{fig:fcs}(a)--\ref{fig:fcs}(c)].
Physically, $W(t)$ counts the particles pumped through the bulk of the chain: the evolution not only lets particles leak into the substrate but also carries those that remain on the chain by $W$ sites toward the left, like a conveyor belt \appref{(Sec.~\ref{sm:sectors} of the Supplemental Material)}.
To detect this pump, we prepare the same initial state and choose a window of $2w$ cells centered on a cut in the middle of the chain [Fig.~\ref{fig:fcs}(a)]; then we switch on $g$ and image the occupations at time $t$.
We keep a snapshot only if every chain site in the window is occupied and every substrate site below it is empty [Figs.~\ref{fig:fcs}(b) and \ref{fig:fcs}(c)]. For each kept snapshot, we record $m=N_L(t)-N_L(0)$, where $N_L$ counts the particles to the left of the cut, in both the chain and the substrate.
The fraction of kept snapshots decays as $\ee^{-4wr_{\rm PBC}(t)}$, so that the post-selection rate itself measures the PBC dynamical free energy on an open chain \appref{(Sec.~\ref{sm:keep} of the Supplemental Material)}.
Each kept snapshot has been refilled by particles carried in across the cut, so that $m$ counts the pumped particles.
Repeating this procedure many times gives the probability $P(m|\mathrm{keep})$ of each $m$ among the kept snapshots, and hence their mean $\langle m\rangle$.
As shown in Fig.~\ref{fig:fcs}(d), this conditional mean approaches $W(t)$ as the window grows.
For example, $P(m=1|\mathrm{keep})=0.991$ for $w=2$ and $N=30$ at $gt=2.5$; the approach is slower for $gt\gtrsim5$, where a winding boundary lies close to the BZ \appref{(Sec.~\ref{sm:mean} of the Supplemental Material)}.
In particular, the jumps of $\langle m\rangle$ reveal the PBC DQPTs on an open chain, although these transitions are absent from $r_{\rm OBC}$.
This measurement thus shows that $W$ is not merely a mathematical winding number, but corresponds to a real physical process.

\paperheading{Summary and discussion}\label{sec:discussion}\label{sec:conclusion}
In summary, we have shown that the subspace DQPTs of a Hermitian chain coupled to a substrate can depend on boundaries even in the thermodynamic limit.
The winding number $W$ decides when they do: open and periodic chains give the same dynamical free energy for $W=0$, even with an NHSE, whereas for $W\neq0$, open boundaries remove the PBC DQPTs and can create a non-Bloch DQPT.
Guided by non-Bloch band theory and confirmed by finite-size numerics, we have found that the OBC dynamical free energy follows from the PBC one, analytically continued along a path with zero winding number.
Moreover, $W$ can be measured by particle counting.
We close with several remarks on the broader implications of these results.

First, in equilibrium, the free energy per site does not depend on boundary conditions in the thermodynamic limit, and it has been natural to study DQPTs under PBC.
Our results show that this expectation can fail out of equilibrium. Unlike in previously reported edge effects on DQPTs~\cite{Khatun2019,Kaur2026,Sedlmayr2018}, a bulk topological invariant decides here whether boundaries matter, a dynamical analog of the bulk-boundary correspondence.

Second, all the non-Hermitian physics here, including the point-gap topology and the NHSE, emerges from a strictly Hermitian system without any approximation, unlike previous studies of non-Hermitian DQPTs~\cite{Zhou2018NonHermitianLatticeDQPT,Qiu2019PTSymmetricQuench,Zhou2021NonHermitianTopologyDQPT,Jing2024BiorthogonalDQPT,Fu2025AnatomyNonHermitianDQPT,Wang2026GaugeInvariantNonHermitianDPT,Jafari2026PseudoHermitianDQPT}.
This places non-Hermitian topology on a microscopic footing and shows that it can arise naturally in subsystems of ordinary closed quantum systems, without engineered gain and loss.

Third, the NHSE, a hallmark of point-gap topology~\cite{Zhang2020Winding,Okuma2020}, does not by itself make DQPTs boundary dependent; this requires a nonzero $W$, which in turn requires both broken time reversal and memory~\footnote{For $\lambda=0$, $H$ is real and symmetric, so that $\mathcal D(-k,\tau)=\mathcal D(k,\tau)$ and $W=0$; a frequency-independent self-energy $\Sigma_{\rm M}(k)$ gives $\mathcal D(k,\tau)=\ee^{-\ii\tau\Tr\Sigma_{\rm M}(k)}\neq0$, without zeros or winding.}.
Memory effects are usually regarded as a complication and removed by the Markov approximation~\cite{Breuer2016,deVega2017}, and DQPTs have been shown to survive them~\cite{Kyaw2020}. Here, they not only occur naturally in a simple substrate model, but also are an essential resource for topology. This suggests engineering non-Markovian environments as a route to new topological phenomena.

Fourth, this topology lives in the time domain: causality forbids such a winding at real frequencies~\footnote{For $\mathrm{Im}\,\omega>0$, $M=\omega-H_S(k)-\Sigma(k,\omega)$ has $(M-M^\dagger)/2\ii\ge\mathrm{Im}\,\omega$, so that $\det M$ neither vanishes nor winds in $k$, and, by the Szeg\H{o}--Widom theorem~\cite{Widom1976,BoettcherSilbermann2006}, $N^{-1}\ln|\det M|$ is then the same under OBC and PBC, also as $\omega$ approaches the real axis.}, so that the boundary dependence has no counterpart in frequency-domain responses and appears only in the real-time dynamics.
Real-time dynamics thus provides a new arena for non-Hermitian topology beyond static spectra.

Fifth, our results extend non-Bloch band theory~\cite{Yao2018,Yokomizo2019,WangAmoeba2024} from static non-Hermitian Hamiltonians to nonequilibrium dynamics with memory, where the non-Hermiticity is generated by the dynamics itself.
The analytic continuation of momentum used here thus provides a practical way to compute open-boundary dynamics in the thermodynamic limit, which may be applied to other nonequilibrium problems.

Finally, our predictions can be tested on several platforms.
The PBC DQPTs may be observed with classical waves, e.g., in waveguide arrays~\cite{Christodoulides2003,Rechtsman2013}, where the substrate replaces engineered gain and loss. Quantum simulators in which DQPTs and dynamical vortices have been observed~\cite{Jurcevic2017,Flaschner2018,Guo2019DQPTQubit,Wang2019PhotonicDQPT} are natural platforms to engineer broken time reversal and memory. Fermionic simulators with site-resolved imaging~\cite{Cheuk2015,Parsons2015} can count the pumped particles and thus measure $W$ directly.
On the theoretical side, the momentum-resolved Green's function of interacting electrons is also a subspace propagator, so that the winding criterion may also govern boundary effects in the dynamics of correlated electrons.

\paperheading{Note added}While completing this work, we became aware of Ref.~\cite{Lin2026}, which shows that the Loschmidt rate functions of SSH chains under OBC and PBC can differ in the thermodynamic limit. It attributes the difference to dynamical zero modes of the open-boundary Loschmidt matrix.
In contrast, the boundary dependence found here is decided by the winding number $W$ and thus has a bulk topological origin, which points to a different mechanism.

\begin{acknowledgments}
We used the AI assistant Claude (Anthropic) to help write, document, and check the numerical code, to prepare figures, and to edit the language of the manuscript. The authors verified all results and take full responsibility for the content.
\end{acknowledgments}

\bibliography{references}

\section*{End Matter}

\begin{figure}[t]
  \centering
  \includegraphics[width=\columnwidth]{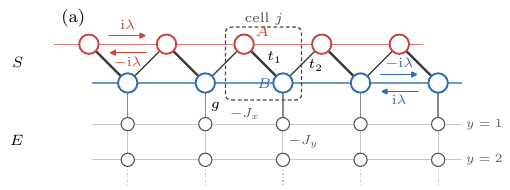}\\[3pt]
  \includegraphics[width=\columnwidth]{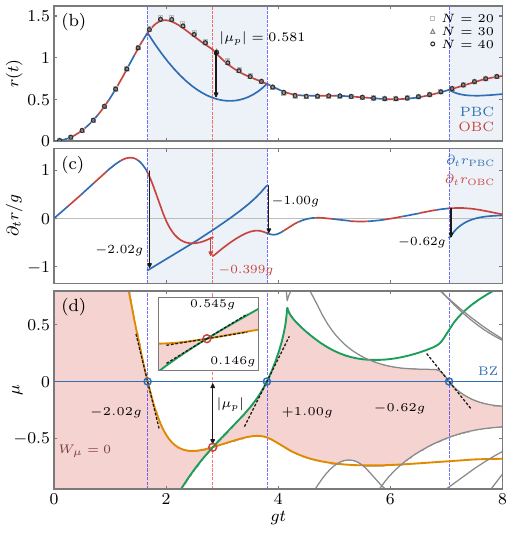}
  \caption{\label{fig:endmatter}\label{fig:hopping}%
    Lattice model and dynamical free energy of open chains.
    (a)~Bonds of $H$, with the amplitude $\langle i|H|j\rangle$ of a hop $j\to i$ along each arrow; dashed box: cell $j$.
    (b)~$r(t)$ for $N\to\infty$ under PBC and OBC [Eqs.~\eqref{eq:rate} and \eqref{eq:robc}], colored as in Fig.~\ref{fig:obc}(a), and for open chains (symbols); shading: $W(t)=1$; arrow: the gap $|\mu_p|$ at $t_p$ [Eq.~\eqref{eq:tp}].
    (c)~Slopes of the $N\to\infty$ curves in (b); arrows: jumps predicted by the tangent slopes in (d).
    (d)~Winding boundaries near the BZ (blue line), colored as in Fig.~\ref{fig:fisher}(a), and the zero-winding plateau (shading), which closes at $(t_p,\mu_p)$ (red circle); dashed: tangents where a winding boundary crosses the BZ (circles) and at $(t_p,\mu_p)$ (inset).
    Vertical lines as in Fig.~\ref{fig:obc}.}
\end{figure}

\appsec{app:model}{Model, parameters, and numerical methods}To specify the model of Eqs.~\eqref{eq:Dk}--\eqref{eq:sigma} on the lattice, we draw the bonds of $H$ in Fig.~\ref{fig:endmatter}(a), where $t_1$, $t_2$, $g$, $-J_x$, and $-J_y$ are real. Under PBC, the Bloch states
\begin{equation}\label{eq:bloch}
|k\alpha\rangle=\frac{1}{\sqrt N}\sum_j\ee^{\ii kj}|j\alpha\rangle,\qquad \alpha=A,B,1,2,\dots,
\end{equation}%
block-diagonalize $H$. In the basis $(A,B,1,2,\dots)$ and with $\varepsilon_k=-2J_x\cos k$, the blocks read
\begin{equation}\label{eq:Hk}
H(k)=\begin{pmatrix}
2\lambda\sin k&t_1+t_2\ee^{-\ii k}&0&0&\cdots\\
t_1+t_2\ee^{\ii k}&-2\lambda\sin k&g&0&\cdots\\
0&g&\varepsilon_k&-J_y&\cdots\\
0&0&-J_y&\varepsilon_k&\ddots\\
\vdots&\vdots&\vdots&\ddots&\ddots
\end{pmatrix}.
\end{equation}
Its upper-left $2\times2$ block is $H_S(k)$ of Eq.~\eqref{eq:hS}, and the rest is the substrate column $H_E(k)$, a semi-infinite chain with the on-site energy $\varepsilon_k$ and the hopping $-J_y$, coupled to $B$ through its first row $|1\rangle$:
\begin{equation}\label{eq:Hblock}
H(k)=\begin{pmatrix}H_S(k)&V\\V^\dagger&H_E(k)\end{pmatrix},\qquad V=g\,|B\rangle\langle1|.
\end{equation}
Eliminating the substrate column (Schur complement) gives the chain block of the resolvent,
\begin{align}
\bigl[\omega-H(k)\bigr]^{-1}_{SS}&=\bigl[\omega-H_S(k)-\Sigma(k,\omega)\bigr]^{-1},\label{eq:schur}\\
\Sigma(k,\omega)&=V\bigl[\omega-H_E(k)\bigr]^{-1}V^\dagger\nonumber\\
&=g^2\,G(k,\omega)\,|B\rangle\langle B|,\label{eq:sigmaG}
\end{align}
where $G(k,\omega)=\langle1|[\omega-H_E(k)]^{-1}|1\rangle$. Removing the first row leaves the same semi-infinite column, so that
\begin{equation}\label{eq:Gself}
G(k,\omega)=\frac{1}{\omega-\varepsilon_k-J_y^2\,G(k,\omega)},
\end{equation}
whose retarded solution, with $G\approx1/\omega$ for large $|\omega|$, is
\begin{equation}\label{eq:Gsol}
G(k,\omega)=\frac{\omega-\varepsilon_k-\sqrt{(\omega-\varepsilon_k)^2-4J_y^2}}{2J_y^2}.
\end{equation}
Inserting it into Eq.~\eqref{eq:sigmaG} gives Eq.~\eqref{eq:sigma}. Finally, for $t>0$,
\begin{equation}\label{eq:Ufourier}
\ee^{-\ii H(k)t}=\ii\int\frac{\mathrm d\omega}{2\pi}\,\ee^{-\ii\omega t}\bigl[\omega+\ii0-H(k)\bigr]^{-1},
\end{equation}
whose chain block is, by Eq.~\eqref{eq:schur}, Eq.~\eqref{eq:Dk}.

\begin{table*}[t]
\caption{\label{tab:params}Parameters and methods of the numerical results. Model: $t_1=g$, $t_2=0.5g$, $\lambda=0.3g$, $J_x=0.15g$, and $J_y=0.25g$. $N$: number of cells of rings and open chains; $L_y$: number of rows to which each substrate column is truncated (Appendix~\ref{app:model}).}
\begin{ruledtabular}
\begin{tabular}{p{0.11\textwidth}p{0.52\textwidth}p{0.31\textwidth}}
Figure & Quantity and method & Parameters\\ \hline
\ref{fig:obc}(a) & $r_{\rm PBC}(t)$: Eq.~\eqref{eq:rate} & $L_y=80$; $2000$ momenta; $\Delta t=0.01/g$\\
\ref{fig:obc}(a) & $r_{\rm OBC}(t)$: Eq.~\eqref{eq:robc}, i.e., $r_{\rm PBC}(\mu_*(t),t)$ along the path $\mu_*(t)$ of Fig.~\ref{fig:fisher}(a) & $L_y=40$; $2048$ momenta (up to $65536$ near $t_p$); $\Delta t=0.01/g$\\
\ref{fig:obc}(a), inset & $\partial_tr$: central differences, taken separately on each side of $t_p$ & $L_y=80$ (PBC; $1000$ momenta), $40$ (OBC); step $0.0025/g$\\
\ref{fig:obc}(b) & PBC curves: zeros $\tau$ of $\mathcal D(k,\tau)$ at real $k$; shading: $W(\tau)$, Eq.~\eqref{eq:Wintro} & $L_y=80$; $600$ momenta\\
\ref{fig:obc}(b) & OBC curves: closure of the zero-winding plateau of $W_\mu(\tau)$ at complex $\tau$ & $L_y=24$; $512$ momenta; grids of $0.01$ in $\mu$ and $0.02/g$ in $\tau$\\
\ref{fig:obc}(b) & circles: zeros of $\det\US(\tau)$; ring from $\prod_n\mathcal D(2\pi n/N,\tau)$, open chain by the argument principle and Newton's method & $N=40$; $L_y=80$ (ring), $24$ (open chain)\\
\ref{fig:phase}(a), (b) & eigenvalues of $\US$: PBC bands; rings and open chains; colors as in Appendix~\ref{app:model} & $N=40$; $L_y=80$ (bands, ring), $24$ (open chain); $2400$ momenta\\
\ref{fig:phase}(c), (d) & $\arg\mathcal D(k,t)$; $W(t)$, Eq.~\eqref{eq:Wintro} & $L_y=80$; $361\times641$ grid in $(k,t)$; $2000$ momenta, $\Delta t=0.01/g$\\
\ref{fig:fisher}(a) & $W_\mu(t)$, Eq.~\eqref{eq:winding_mu}; $\mu_*(t)=-0.8136$ for $gt\le2$, $-0.2536$ for $gt\ge4$, and linear in between; dots: $r_{\rm PBC}(t)-r_{\rm OBC}(t)$ from the curves of Fig.~\ref{fig:obc}(a); orange and green: tracked zeros, as for Fig.~\ref{fig:endmatter}(d) & $L_y=40$ (map), $24$ (orange, green); $384$ momenta; grids of $0.01$ in $\mu$ and $0.02/g$ in $t$; dots every $0.2/g$\\
\ref{fig:fisher}(b), (c) & $\ee^{-\ii Ht}|r\rangle$ on an open chain & $N=40$; $L_y=12$; $\Delta t=0.002/g$\\
\ref{fig:fcs} & counting statistics, Eq.~\eqref{eq:fcs} & $N=30$; $L_y=24$; $32$ counting fields; $\Delta t=0.02/g$\\
\ref{fig:endmatter}(b) & $r_N(t)$ of open chains & $N=20$, $30$, $40$; $L_y=24$; $\Delta t=0.05/g$\\
\ref{fig:endmatter}(c) & $\partial_tr_{\rm PBC}$ and $\partial_tr_{\rm OBC}$ from the curves of Fig.~\ref{fig:obc}(a), by central differences on each side of the kinks; measured slope jumps: quadratic fits on each side of $t_c$, over $0.03<g|t-t_c|<0.15$ (PBC) and $0.004<g|t-t_p|<0.06$ (OBC); arrows: jumps predicted by the tangent slopes in (d) & as for Fig.~\ref{fig:obc}(a)\\
\ref{fig:endmatter}(d) & winding boundaries: zeros of $\mathcal D$ located by the winding of $\arg\mathcal D$ around the cells of a polar grid and refined by Newton's method; tangent slopes $\dot\mu_n$: quadratic fits through $8$ consecutive times; shading: $W_\mu=0$, as for Fig.~\ref{fig:fisher}(a) & $L_y=24$ (lines), $40$ (shading); $256$ momenta; grid of $0.01$ in $\mu\in[-1.4,1]$; $\Delta t=0.01/g$\\
\ref{fig:endmatter}(d), inset & plateau edges $\mu_\pm(t)$: intersections of the linear pieces of $r_{\rm PBC}(\mu,t)$ [Eq.~\eqref{eq:max}] on three circles below, on, and above the plateau; tangent slopes: quadratic fits through the $8$ times with $g|t-t_p|<0.025$ & $L_y=24$; $4096$ momenta\\
\end{tabular}
\end{ruledtabular}
\end{table*}%
At complex momenta, $H(k)$ is not Hermitian, and we evaluate $\ee^{-\ii H(k)t}$ through its eigendecomposition $H=R\Lambda R^{-1}$. To evaluate $\US$ at real and complex momenta and on open chains, we truncate each substrate column to $L_y$ rows; open chains have $N$ cells on $N$ substrate columns. The truncation does not matter as long as a particle that leaves the chain cannot reach the last row. Since the group velocity in the substrate is at most $2J_y$, this takes at least the time $L_y/(2J_y)$, which exceeds $11/g$ for $L_y>L_c=22J_y/g=5.5$; for $gt\le11$, we use $L_y\ge12$, well beyond $L_c$. Table~\ref{tab:params} lists all parameters and methods.

To color the eigenvalues of open chains in Fig.~\ref{fig:phase} by where their eigenvectors live, we use the weights $p_j$ of each normalized right eigenvector $\psi$ of $\US$ on the cells,
\begin{equation}\label{eq:ipr}
p_j=\sum\nolimits_\alpha|\psi_{j\alpha}|^2,\qquad\bar j=\sum\nolimits_jjp_j,\qquad\mathrm{IPR}=\sum\nolimits_jp_j^2:
\end{equation}
the color is red (blue) for $\bar j<(N+1)/2$ [$\bar j>(N+1)/2$], with the depth $\min(\mathrm{IPR}/0.25,1)$.

\appsec{app:nonbloch}{Dynamical free energy of open chains}To check Eq.~\eqref{eq:robc} against finite-size numerics, Fig.~\ref{fig:endmatter}(b) compares it with open chains of $N=20$, $30$, and $40$ cells (Table~\ref{tab:params}). They approach it as $1/N$ and, extrapolated by a least-squares fit $r_N=a+b/N$ at each time, agree with it to within $1.0\times10^{-3}$, except for $|g(t-t_p)|<0.45$, where finite chains round the kink. We do not use larger $N$, since for $W=1$ the smallest singular value of $\US$ decreases exponentially with $N$ and, at $gt=2.5$, reaches double precision near $N=55$.

In the main text, we stated that the slope jumps of all kinks follow from the tangent slopes of the winding boundaries. Figures~\ref{fig:endmatter}(c) and \ref{fig:endmatter}(d) check this numerically. Where a winding boundary crosses the BZ [circles in Fig.~\ref{fig:endmatter}(d)], its tangent slope $\dot\mu_n=-2.02g$, $1.00g$, and $-0.62g$ at $gt_c=1.6695$, $3.7976$, and $7.0472$ predicts the jump $-|\dot\mu_n|$ of the slope of $r_{\rm PBC}$; the measured jumps are $-2.06g$, $-1.00g$, and $-0.61g$ [Fig.~\ref{fig:endmatter}(c)]. At $gt_p=2.8306$, where the plateau closes at $\mu_p=-0.5811$, $r_{\rm PBC}$ is smooth, whereas the upper edge $\mu_+(t)$ in Eq.~\eqref{eq:tp} switches from the orange winding boundary, with the slope $0.146g$, to the green one, with the slope $0.545g$ [inset of Fig.~\ref{fig:endmatter}(d)]. The predicted jump of the slope of $r_{\rm OBC}$, $-(0.545-0.146)g=-0.399g$, agrees with the measured $-0.3986g$ [Fig.~\ref{fig:endmatter}(c)].

\clearpage
\onecolumngrid
\setcounter{topnumber}{1}%
\setcounter{secnumdepth}{1}
\setcounter{equation}{0}\setcounter{figure}{0}\setcounter{table}{0}\setcounter{section}{0}
\renewcommand{\theequation}{S\arabic{equation}}\renewcommand{\theHequation}{S\arabic{equation}}
\renewcommand{\thefigure}{S\arabic{figure}}\renewcommand{\theHfigure}{S\arabic{figure}}
\renewcommand{\thetable}{S\arabic{table}}\renewcommand{\theHtable}{S\arabic{table}}
\renewcommand{\thesection}{S\arabic{section}}\renewcommand{\theHsection}{S\arabic{section}}
\begin{center}
{\large\bfseries Supplemental Material for ``Non-Bloch Dynamical Quantum Phase Transitions with Memory''}
\end{center}

\noindent This Supplemental Material shows, with exact numerics, why the counting protocol of Fig.~\ref{fig:fcs} measures $W(t)$; Sec.~\ref{sm:method} explains how the numbers are computed. Throughout, $g=1$, the parameters are those of the main text, and the open chains have $N=30$ cells on $L_y=24$ substrate rows; $r_{\rm PBC}$, $r_{\rm OBC}$, and $W(t)$ are those of Table~\ref{tab:params}, and the phase of $\mathcal D$ in Fig.~\ref{fig:smmdist}(a) uses $L_y=24$ and $360$ momenta.

\section{Probability of a snapshot}\label{sm:setup}

\begin{figure}[h]
  \begin{minipage}[c]{0.5\textwidth}
    \includegraphics[width=\linewidth]{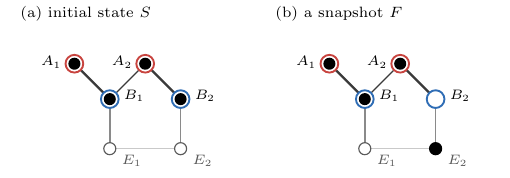}
  \end{minipage}\hfill
  \begin{minipage}[c]{0.46\textwidth}
  \caption{\label{fig:smsnapshot}%
    A two-cell lattice with one substrate site $E_j$ below each $B_j$, for illustration. (a)~Initial state: the chain sites $S$ are filled and the substrate is empty. (b)~A snapshot $F$, whose probability is Eq.~\eqref{eq:slater}.}
  \end{minipage}
\end{figure}

Since the fermions do not interact, each particle evolves on its own with $U=\ee^{-\ii Ht}$, and the amplitude of a snapshot is a determinant of one-particle amplitudes. For the two-cell lattice of Fig.~\ref{fig:smsnapshot}, with the initial particles $S=\{A_1,B_1,A_2,B_2\}$ and the snapshot $F=\{A_1,B_1,A_2,E_2\}$,
\begin{equation}\label{eq:slater}
P(F)=\bigl|\langle F|\Psi(t)\rangle\bigr|^2=\bigl|\det U[F,S]\bigr|^2
=\left|\det\begin{pmatrix}
U_{A_1A_1}&U_{A_1B_1}&U_{A_1A_2}&U_{A_1B_2}\\
U_{B_1A_1}&U_{B_1B_1}&U_{B_1A_2}&U_{B_1B_2}\\
U_{A_2A_1}&U_{A_2B_1}&U_{A_2A_2}&U_{A_2B_2}\\
U_{E_2A_1}&U_{E_2B_1}&U_{E_2A_2}&U_{E_2B_2}
\end{pmatrix}\right|^2,
\end{equation}
with rows for the occupied sites $F$ and columns for the initial particles $S$; the determinant adds up all ways of sending the four particles to $F$.

Every probability below is a sum of Eq.~\eqref{eq:slater} over snapshots; Sec.~\ref{sm:method} explains how we evaluate these sums exactly.

\section{How often a snapshot is kept}\label{sm:keep}

\begin{figure}[t]
  \centering
  \includegraphics[width=0.85\textwidth]{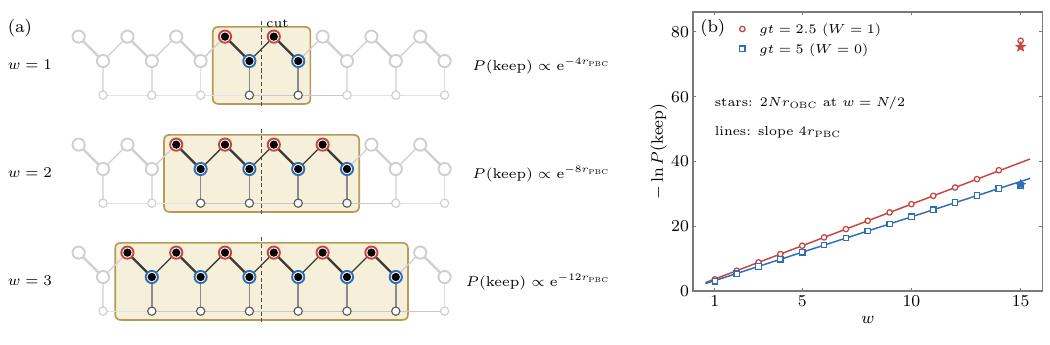}
  \caption{\label{fig:smkeep}%
    Kept snapshots and their probability. (a)~Windows of $w=1$, $2$, and $3$ cells on each side of the cut (dashed); a snapshot is kept if the window (frame) is back in its initial pattern, whatever happens elsewhere (gray; one substrate row shown). (b)~Exact post-selection probability versus $w$ at $gt=2.5$ ($W=1$) and $gt=5$ ($W=0$) (symbols); lines: slope $4r_{\rm PBC}$ [Eq.~\eqref{eq:Pkeep}]; stars: $2Nr_{\rm OBC}$ at $w=N/2$, where the window covers the whole open chain.}
\end{figure}

The probability of keeping a snapshot decays exponentially with the window size [Fig.~\ref{fig:smkeep}],
\begin{equation}\label{eq:Pkeep}
P(\mathrm{keep})\propto\ee^{-4wr_{\rm PBC}(t)},
\end{equation}
with the fitted slopes $2.565$ at $gt=2.5$ and $2.188$ at $gt=5$, against $4r_{\rm PBC}=2.564$ and $2.183$.

To see why, consider first a ring of $N$ cells and a window that covers the whole ring. A snapshot is then kept only if all particles are back on the ring, so that its probability is the Loschmidt echo, which defines $r_{\rm PBC}$ [Eqs.~\eqref{eq:subdqpt_rate} and \eqref{eq:rate}]:
\begin{equation}\label{eq:Pring}
P(\mathrm{keep})\big|_{\rm ring}=\bigl|\det\US^{\rm ring}\bigr|^2=\prod_k\bigl|\mathcal D(k,t)\bigr|^2\approx\ee^{-2Nr_{\rm PBC}(t)}.
\end{equation}
Each cell thus contributes the factor $\ee^{-2r_{\rm PBC}}$. Although our chain is open, a window of $2w$ cells far from its ends behaves as a part of a ring, so that replacing $N$ by $2w$ in Eq.~\eqref{eq:Pring} gives Eq.~\eqref{eq:Pkeep}: the number of cells in the window sets how much dynamical free energy enters its return probability. Only a window that reaches the ends of the open chain, $w=N/2$, feels them, and $-\ln P(\mathrm{keep})$ then jumps to $2Nr_{\rm OBC}$ [stars in Fig.~\ref{fig:smkeep}(b): $77.1$ against $75.3$ at $gt=2.5$, and $32.5$ against $32.7$ at $gt=5$].

\section{Distribution of the transferred charge}\label{sm:sectors}

\begin{figure}[h]
  \centering
  \includegraphics[width=0.85\textwidth]{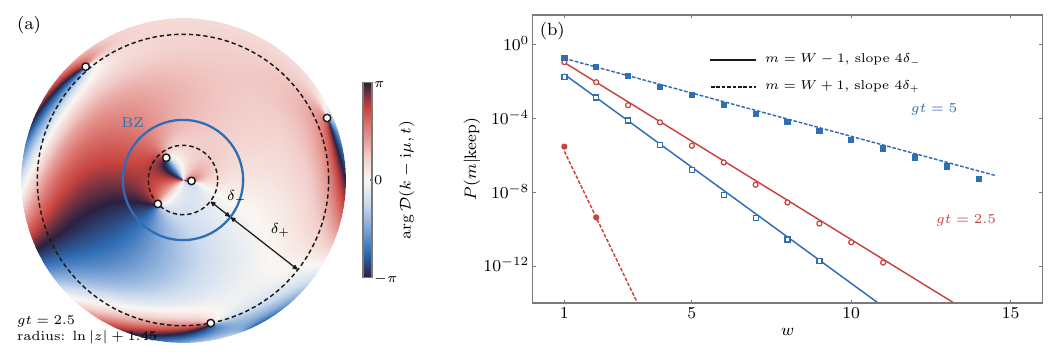}
  \caption{\label{fig:smmdist}%
    Cost of the wrong charges. (a)~Phase of $\mathcal D(k-\ii\mu,t)$ in the plane of $z=\ee^{\mu+\ii k}$ at $gt=2.5$ ($W=1$), with the radial coordinate $\ln|z|+1.45$; white circles: zeros of $\mathcal D$; blue: the BZ, $|z|=1$; dashed: the circles through the nearest zeros inside and outside the BZ, at the distances $\delta_-=0.613$ and $\delta_+=2.064$ in $\ln|z|$ [Eq.~\eqref{eq:Pm}]. (b)~Exact $P(m=W\mp1|\mathrm{keep})$ versus $w$ at $gt=2.5$ (red) and $gt=5$ (blue) (symbols, values above $10^{-12}$; open: $m=W-1$, filled: $m=W+1$); lines: slopes $4\delta_\mp$ [Eq.~\eqref{eq:Pm}].}
\end{figure}

Among the kept snapshots, the charges $m=W\pm1$ are exponentially rare [Fig.~\ref{fig:smmdist}(b)],
\begin{equation}\label{eq:Pm}
P(m=W\pm1|\mathrm{keep})\propto\ee^{-4w\delta_\pm},
\end{equation}
where $\delta_+$ ($\delta_-$) is the distance from the BZ to the nearest winding boundary above (below) it, i.e., the distance in $\ln|z|$ from the unit circle to the nearest zero of $\mathcal D$ outside (inside) it [Fig.~\ref{fig:smmdist}(a)]. The fitted slopes are $2.48$ at $gt=2.5$, against $4\delta_-=2.45$, and $2.95$ and $1.13$ at $gt=5$, against $4\delta_-=2.90$ and $4\delta_+=1.08$. At $gt=2.5$, $P(m=W+1|\mathrm{keep})$ falls so fast, with $4\delta_+=8.25$, that it reaches the double-precision floor of about $10^{-13}$ already at $w=3$. Charges with $|m-W|\ge2$ are rarer still [Fig.~\ref{fig:smmean}(a)]. A kept snapshot thus has $W$ particles carried across the cut, as by a conveyor belt that moves the particles on the chain by $W$ sites.

\section{Mean transferred charge}\label{sm:mean}

\begin{figure}[t]
  \centering
  \includegraphics[width=0.85\textwidth]{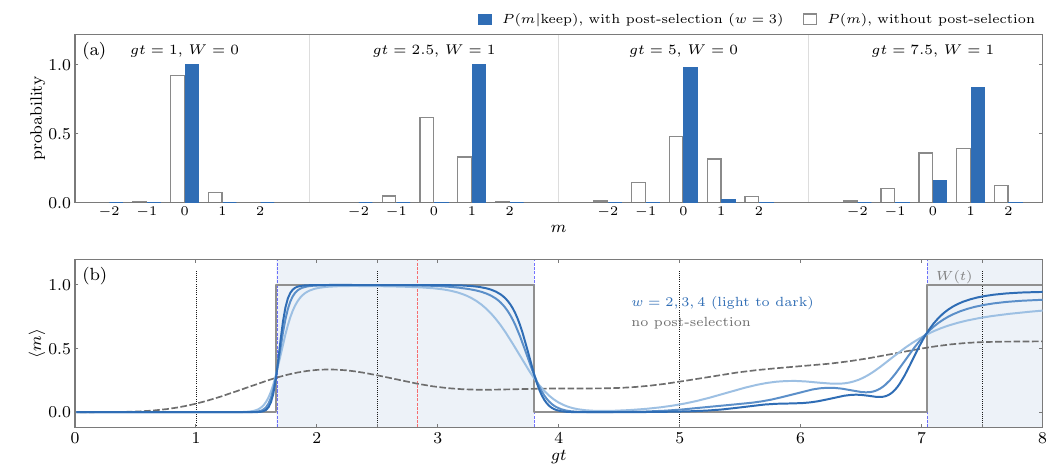}
  \caption{\label{fig:smmean}%
    Distribution and mean of the transferred charge. (a)~$P(m|\mathrm{keep})$ for $w=3$ (filled) and $P(m)$ without post-selection (open) at $gt=1$, $2.5$, $5$, and $7.5$ [dotted lines in (b)]. (b)~$\langle m\rangle$ for $w=2$, $3$, and $4$ (light to dark blue) and without post-selection (dashed), compared with $W(t)$ (gray); shading: $W(t)=1$; dashed lines: PBC DQPTs (blue) and $t_p$ (red).}
\end{figure}

For a large window far from the ends of the chain, $1\ll w\ll N$, all charges other than $W$ die out [Eq.~\eqref{eq:Pm}], so that the kept snapshots have the mean charge
\begin{equation}\label{eq:mean}
\langle m\rangle\approx W+P(m=W+1|\mathrm{keep})-P(m=W-1|\mathrm{keep})\ \to\ W.
\end{equation}
Numerically, $P(m=W|\mathrm{keep})$ is $0.991$, $0.9995$, and $0.9999$ for $w=2$, $3$, and $4$ at $gt=2.5$; for $w=3$, it is $1.000$, $0.999$, $0.980$, and $0.838$ at $gt=1$, $2.5$, $5$, and $7.5$ [Fig.~\ref{fig:smmean}(a)]. The approach to $W(t)$ is slow only where a winding boundary lies close to the BZ, i.e., where $\delta_+$ or $\delta_-$ is small, as near the PBC DQPTs and at $gt=5$, where $\delta_+=0.27$ [Figs.~\ref{fig:smmean}(b) and \ref{fig:fcs}(d)]. Without post-selection, $m$ is broadly distributed, and $\langle m\rangle=0.290$ at $gt=2.5$ is not quantized.

\section{How we compute the counting statistics}\label{sm:method}

All probabilities in Fig.~\ref{fig:fcs}(d) and Figs.~\ref{fig:smkeep}--\ref{fig:smmean} are sums of Eq.~\eqref{eq:slater} over snapshots; for example, $P(\mathrm{keep},m)$ is the sum over the kept snapshots with the charge $m$. These sums cannot be done snapshot by snapshot: on the open chain of $N=30$ cells and $L_y=24$ rows, the $60$ particles occupy $780$ sites in $\binom{780}{60}\approx4\times10^{90}$ ways. Sampling snapshots does not help either, since the probabilities of interest are as small as $10^{-12}$ [Fig.~\ref{fig:smmdist}(b)]. We therefore use the counting field $\chi$ of full counting statistics~\cite{Levitov1996,Klich2003}, a real parameter conjugate to $m$, and compute the generating function
\begin{equation}\label{eq:Zdef}
\mathcal Z(\chi)=\sum_mP(\mathrm{keep},m)\,\ee^{\ii\chi m}.
\end{equation}

\begin{figure}[h]
  \centering
  \includegraphics[width=0.7\textwidth]{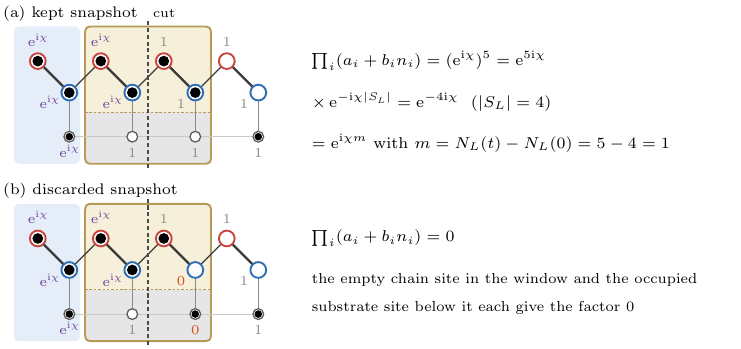}
  \caption{\label{fig:smweights}%
    Weights of Table~\ref{tab:weights} for two snapshots of a lattice of four cells with $w=1$ (one substrate row shown; dashed: cut). The chain starts filled, so that $|S_L|=4$. Each site carries its factor $a_i+b_in_i$ (gray: $1$, red: $0$, purple: $\ee^{\ii\chi}$); black dots: occupied sites; cream (gray) shading: chain (substrate) sites of the window; blue shading: the other sites left of the cut. (a)~A kept snapshot, with five particles left of the cut, has the weight $\ee^{\ii\chi m}$ with $m=1$. (b)~A discarded snapshot has the weight $0$.}
\end{figure}

The post-selection and the counting enter $\mathcal Z(\chi)$ through one weight per site. A snapshot is a list of occupations $n_i\in\{0,1\}$ of all sites $i$, and we give it the weight $\prod_i(a_i+b_in_i)$, i.e., each site contributes $a_i$ if it is empty and $a_i+b_i$ if it is occupied, with the values in Table~\ref{tab:weights}. A snapshot with an empty chain site in the window or an occupied substrate site below the window thus gets the weight zero, i.e., it is discarded, and in a kept snapshot each particle left of the cut contributes the factor $\ee^{\ii\chi}$:
\begin{equation}\label{eq:weight}
\prod_i\bigl(a_i+b_in_i\bigr)=\begin{cases}0 & \text{for a discarded snapshot},\\ \ee^{\ii\chi N_L(t)} & \text{for a kept snapshot},\end{cases}
\end{equation}
where $N_L(t)$ is the number of particles left of the cut. Averaging Eq.~\eqref{eq:weight} over all snapshots and multiplying by $\ee^{-\ii\chi|S_L|}$, where $|S_L|=N_L(0)$ is the number of particles that start left of the cut, gives Eq.~\eqref{eq:Zdef}, since $m=N_L(t)-N_L(0)$. Figure~\ref{fig:smweights} works this out for two snapshots of a small lattice.

\begin{table}[h]
\caption{\label{tab:weights}Weights of the sites in Eq.~\eqref{eq:weight}.}
\begin{ruledtabular}
\begin{tabular}{lccc}
Sites & empty, $a_i$ & occupied, $a_i+b_i$ & $(a_i,b_i)$\\
\hline
chain sites in the window, left of the cut & $0$ & $\ee^{\ii\chi}$ & $(0,\ee^{\ii\chi})$\\
chain sites in the window, right of the cut & $0$ & $1$ & $(0,1)$\\
substrate sites below the window & $1$ & $0$ & $(1,-1)$\\
other sites left of the cut & $1$ & $\ee^{\ii\chi}$ & $(1,\ee^{\ii\chi}-1)$\\
other sites right of the cut & $1$ & $1$ & $(1,0)$
\end{tabular}
\end{ruledtabular}
\end{table}

For noninteracting fermions, the state remains a Slater determinant, and the average of a product of site weights is a single determinant~\cite{Klich2003}:
\begin{equation}\label{eq:fcs}
\mathcal Z(\chi)=\Bigl\langle\prod_i\bigl(a_i+b_i\hat n_i\bigr)\Bigr\rangle_t\,\ee^{-\ii\chi|S_L|}=\det\bigl[a+b\,C(t)\bigr]\,\ee^{-\ii\chi|S_L|}.
\end{equation}
Here, $a$ and $b$ are the diagonal matrices with the entries $a_i$ and $b_i$, and $C(t)$ is the one-particle correlation matrix. Since the particles start on the chain sites $s\in S$ and evolve with $U=\ee^{-\ii Ht}$,
\begin{equation}\label{eq:corr}
C_{ij}(t)=\langle c_i^\dagger c_j\rangle_t=\sum_{s\in S}U_{is}^*\,U_{js}.
\end{equation}
A discrete Fourier transform over $32$ values of $\chi$ then gives
\begin{equation}\label{eq:Pinv}
P(\mathrm{keep},m)=\frac{1}{32}\sum_{l=0}^{31}\mathcal Z\Bigl(\frac{2\pi l}{32}\Bigr)\,\ee^{-2\pi\ii lm/32},
\end{equation}
which resolves all charges with $|m|<16$; the charges with $|m|\le3$ already carry all but $4\times10^{-4}$ of the probability for $gt\le8$ (less than $10^{-8}$ among the kept snapshots). From $P(\mathrm{keep},m)$, we obtain $P(\mathrm{keep})=\mathcal Z(0)$, $P(m|\mathrm{keep})=P(\mathrm{keep},m)/P(\mathrm{keep})$, and $\langle m\rangle=\sum_mm\,P(m|\mathrm{keep})$. Without post-selection, only the counting weights remain, and Eq.~\eqref{eq:fcs} reduces to the standard result $\mathcal Z(\chi)=\det[1+(\ee^{\ii\chi}-1)\,C_L(t)]\,\ee^{-\ii\chi|S_L|}$~\cite{Levitov1996,Klich2003}, where $C_L$ is the block of $C(t)$ on the sites left of the cut [open bars in Fig.~\ref{fig:smmean}(a) and dashed line in Fig.~\ref{fig:smmean}(b)].

In practice, we diagonalize the $780\times780$ matrix $H$ once and evaluate Eq.~\eqref{eq:corr} at each time. A site with $(a_i,b_i)=(1,0)$ contributes a row of the identity matrix to $a+bC$ and drops out of the determinant, which thus runs only over the sites left of the cut or in the window, e.g., $468$ sites for $w=3$. Two checks confirm the numerics. For $w=N/2$, the window covers the whole chain, and $-\ln P(\mathrm{keep})$ agrees with $-\ln|\det\US|^2$ of the open chain [Fig.~\ref{fig:endmatter}(b)] to all printed digits, e.g., $77.1276$ at $gt=2.5$. Substrates of $L_y=12$ and $24$ rows give the same results. The slopes quoted in Secs.~\ref{sm:keep} and \ref{sm:sectors} are least-squares fits over $w=2$--$10$; for $P(m|\mathrm{keep})$, we use only values above $10^{-12}$, since Eq.~\eqref{eq:Pinv} resolves $P(m|\mathrm{keep})$ only down to about $10^{-13}$.

\end{document}